# Restate the reference for EEG microstate analysis

Shiang Hu[1], Esin Karahan[1], Pedro A. Valdes-Sosa[1,2*]

[1]The clinical hospital of Chengdu Brain Science Institute, MOE key lab for NeuroInformation, University of Electronic Science and Technology of China, Chengdu, China

[2]Cuban Neuroscience Center, Havana, Cuba

Email: pedro.valdes@neuroinformatics-collaboratory.org

## Abstract

Despite the decades of efforts, the choice of EEG reference is still a debated fundamental issue. Non-neutral reference can inevitably inject the uncontrolled temporal biases into all EEG recordings, which may influence the spatiotemporal analysis of brain activity. A method, termed *microstates*, identifying spatiotemporal EEG features as the quasi-stable topography states in milliseconds, suggests its potential as biomarkers of neurophysiological disease. As reference electrode standardization technique (REST) could reconstruct an infinity reference approximately, it is a question whether REST or the other references will be more reliable than average reference (AR) for the microstates analysis. In this study, we design the microstate-based EEG forward model, and apply different references for microstates analysis. The spatial similarity between the generated and assumed cluster maps is mainly investigated. Furthermore, the real EEG data by the parametric bootstrap method is used to validate the performance of the references. Finally, we find that REST is robust to recover more similar cluster maps to the assumption than AR in the simulation, and the cluster maps between REST and AR on the real EEG data are quite different. This study may indicate that REST contributes to identifying more objective microstates features than AR.

## 1. Introduction

The choice of electroencephalograph (EEG) reference is an unresolved fundamental issue in clinical neurophysiology. Though many references have been proposed, there is still no a clear-cut demonstration of the superiority of one reference over another until the recently proposed unified reference framework (Hu et al., 2018). We will attempt to shed light on this discussion by comparing the effect of selected reference on the segmentation of EEG into microstates, referred to by D. Lehmann as the 'atoms of thought' (Lehmann et al., 1998; Michel and Koenig, 2017).

The reference problem exists since an EEG recording is the difference of potentials measured at two electrodes, namely, active electrode and reference electrode. The active electrode is placed on the scalp, and ideally it should only measure the sources from a specific brain region. Reference electrode could be put on the scalp, body, or virtually the linear combination of all the electrodes, and it theoretically should neither be affected by other cerebral sources nor pick up unwanted brain activity. A naive approach might be to place it very far from the head. However, this is not practical since the reference would become an antenna for undesired environmental fields. If there were no such fields, one would be recording with an ideal (but practically unattainable) 'infinite reference' (IR).

A poor man's replacement for the IR is to choose a 'physical body reference' (PBR) that would allow cancelation of the undesired environmental fields by the EEG differential amplifier and hopefully not record much brain activity. Proposals for such a PBR have been the vertex, linked ears, the tip of nose, etc. (Teplan, 2002). Besides the monopolar PBR reference, it can be bipolar reference as well, an example of which is ipsilateral mastoids (or ears), being frequently adopted in clinical practice. Unfortunately, all PBRs are doomed to fail since there is no point on the scalp or body surface where the potential is zero or constant (Nunez and Srinivasan, 2006).

In view of the shortcomings of PBRs, efforts have turned to obtaining a virtual reference as a mathematical transformation of the recorded EEG. The best known virtual reference is the average reference (AR) (Lehmann, 1971; Offner, 1950). The rationale behind this proposal is that EEG potentials recorded from a dense electrode array placed on a closed surface nearly sum to zero (Bertrand et al., 1985; Nunez et al., 1997). The advantage of the AR only comes into playing with a large number of electrodes and extensive coverage of the head (Christodoulakis et al., 2013). These requirements are never fulfilled. This explains the mediocre performance in some cases of AR as highlighted in the literature (Bertrand et al., 1985; Dien, 1998).

An alternative and biophysically motivated virtual reference is 'reference electrode standardization technique' (REST) (Yao, 2001). It consists of finding a source configuration that explains observed EEG, and then projecting the sources back to the electrodes again—but first eliminating the effect of the reference. Thus, REST is a theoretical reconstruction of what the EEG would be like if one were using an ideal and noiseless IR. One could be worried that there is no unique source configuration that explains the voltages since the EEG inverse problem is ill-posed. However, it should be obvious that REST does not depend on which inverse solution is used. There are some evidences that REST provides better estimates of spectral mapping (Yao et al., 2005), coherency (Marzetti et al., 2007), spatiotemporal analysis of evoked potentials (Yao et al., 2007), default model network (Qin et al., 2010), scalp EEG potentials (Liu et al., 2015), connectivity (Chella et al., 2016) and so on.

In fact, there seem to be two opposing views on the issue of the reference:

- Proponents of the AR argue that the reference is irrelevant for topographic analysis. In their view, an EEG topography is analogous to a configuration of peaks and valleys. The only effect of a reference would be to change the "sea level" which does not change the overall landscape (Geselowitz, 1998; Michel et al., 2004). Thus, the choice of a 'best' reference is not essential and the AR would be the simplest choice.
- A contrasting opinion is that the correct reference is essential for spatiotemporal model estimation. The consequence of a wrong reference is equivalent to adding an arbitrary and (possibly very structured time series) to all recording, which could certainly bias dynamical parameter estimation. An example of this type of bias has been shown by Marzetti et al. (Marzetti et al., 2007), who showed that the use of AR did not allow the optimal reconstruction of simulated networks from EEG coherence measures. This is not surprising since coherency is obtained from a Fourier analysis of the EEG and is certainly not invariant to the addition of an arbitrary signal.

Thus, superficially it might seem that there is clear 'division of labor' for references, in which topographically oriented questions should be answered with the AR, and dynamical oriented questions would benefit best using REST. However, there are problems that are usually considered topography for which dynamical modeling is also essential. Evaluating the effect of reference in such situations could be a crucial test for reference estimation procedures.

A 'canonical' instance of dynamical modeling of topographies is the parsing of a reference corrected multichannel EEG recording into a sequence of *microstates*. Each microstate is a consecutive set of EEG topographies, lasting up hundreds of milliseconds, which differ only in strength and polarity. The commonly-used steps of microstates analysis are to submit topographic maps into a clustering algorithm that is based upon a distance metric between maps designed to be insensitive to strength and polarity, and that presupposes the correction of the AR (Khanna et al., 2015; Murray et al., 2008). Each resulting clustering centroids or 'cluster map' is considered as the topographic map of a given microstate. The classification scheme partitions the space of all the topographic maps based on the metric by means of nearest neighbor criteria. The entire EEG signals will thus be segmented into an alternating series of cluster maps active over the discrete time intervals. Microstates have been found to be reliably identifiable across subjects, to vary consistently over the lifespan, and seem to be the valid biomarkers of brain disorders (Kikuchi et al., 2011; Lehmann et al., 2005; Nishida et al., 2013). In view of this literature, we ask ourselves the question: is microstates analysis significantly affected by the choice of a different reference?

To answer this question, we designed a forward model to simulate the resting state EEG by a sequence of underlying microstates. In the simulation, different microstates appeared at random, each occurrence lasting a variable duration. Strength and polarity modifications of the basic cluster maps could vary smoothly for the duration of that state. Subsequently, microstates analysis was performed on the simulated EEG but introducing both AR and REST into the definition of distance metric for clustering and classification. We show that the choice of reference leads to statistically significant microstates analysis. This result was confirmed by comparing the microstates analysis of the real EEG data via resampling methods.

## 2. EEG reference schemes

In what follows, we denote scalars with lowercase symbols (e.g. $x$), vectors with lowercase bold (e.g. $\mathbf{x}$), matrices with uppercase bold (e.g. $\mathbf{X}$); unknown parameters will be denoted by Greek letters (e.g. $\xi$). Furthermore, $\mathbf{1}$ is the vector of ones; $\mathbf{I}_x$ is a $x$ by $x$ identity matrix; $N(\boldsymbol{\mu}, \boldsymbol{\Sigma})$ is the multivariable Gaussian distribution with mean vector $\boldsymbol{\mu}$ and covariance matrix $\boldsymbol{\Sigma}$; $(\cdot)^\mathbf{T}$ is the transpose of $(\cdot)$; $\mathbf{X}^+$ is the pseudo-inverse of $\mathbf{X}$; $\hat{\mathbf{x}}$ is the estimation of $\mathbf{X}$; $\|\cdot\|_2$ is the $l_2$ norm; We state that $N_e$ is number of electrodes, $N_v$ is the number of brain sources, $N_\mu$ is the number of templates, $N_t$ is the number of templates; $E\{\cdot\}$ and $C\{\cdot\}$ denote the Expectation and the Covariance operator, respectively; the symbol $\otimes$ represent the Kronecker product.

EEG reference schemes can be divided into two categories, that is, online references, and offline re-references. Online reference is exactly the recording reference put on the body (i.e. PBR); and offline re-reference is to minimize the shortcomings of PBR. Linked mastoids, AR, and REST are three canonical re-references. Among the three, linked mastoids has been criticized for a long time due to the 'shunting' phenomenon (Garneski and Steelman, 1958; Kaiser, 2000; McAvoy and Little, 1949; Nunez, 1991) and the distortion of power and coherence spectra (Chella et al., 2016; Marzetti et al., 2007; Shaw, 1984; Travis, 1994). Thus, linked-mastoids is not discussed here.

### 2.1. Physical Body Reference (PBR)

Without loss of generality, the seemingly simplest and commonly-used PBR, FCz electrode, is to serve as the recording reference. We assume that the recorded EEG potentials are measured as the difference between the electric potentials of the active electrodes and that of the PBR electrode (FCz), which can be shown as,

$$\mathbf{v}^{PBR} = (\mathbf{I}_{N_e} - \mathbf{1}\mathbf{f}^\mathbf{T}) \cdot \mathbf{v}^{IR} \tag{1}$$

where $\mathbf{v}^{PBR}$ and $\mathbf{v}^{IR}$ are the measured EEG potentials over $N_e$ electrodes at an instant with PBR (here, FCz electrode) and IR, respectively; $\mathbf{f}^{T} = [0,...0,1,0,...0]$ is a zeros vector except for only one entry being 1 at the corresponding channel of FCz.

## 2.2. Average reference (AR)

If the PBR is monopolar, AR is easily conducted by subtracting the mean of potentials from all electrodes at each time sample. It is formulated as

$$\mathbf{v}^{AR} = (\mathbf{I}_{N_e} - \mathbf{1}\mathbf{f}_1^{T})\mathbf{v}^{PBR} \tag{2}$$

where $\mathbf{v}^{AR}$ is the EEG potentials with AR, $\mathbf{f}_1^{T} = [1/N_e,...,1/N_e]$ is a vector full of $1/N_e$.

## 2.3. Reference electrode standardization technique (REST)

Exploiting the fact that source activities are reference independent, REST aims to reestablish a virtual reference at infinity from the PBR or the linear combination of some electrodes (e.g. linked mastoids, AR) approximately (Yao, 2001; Yao et al., 2005).

The IR is the desired neutral reference which does not exist in the human body. However, the scalp EEG potentials with IR are theoretically existing due to the discretized approximation of Maxell equation

$$\mathbf{v}^{IR} = \mathbf{G} \cdot \mathbf{s} \tag{3}$$

where $\mathbf{G}$ is the lead field referenced with IR with the size $N_e \times N_v$, only dependent on the head model, source configuration and electrode montage (Hu et al., 2017); $\mathbf{s}$ is the primal current density of $N_v$ equivalent sources; measurement noise is assumed to be zero in this model.

Similarly, we have

$$\mathbf{v}^{AR} = \mathbf{G}^{AR} \cdot \mathbf{s}, \ \mathbf{G}^{AR} = (\mathbf{I}_{N_e} - \mathbf{1}\mathbf{f}_1^{T}) \cdot \mathbf{G} \tag{4}$$

where $\mathbf{G}^{AR}$ is the lead field referenced with AR.

The primal current density $\mathbf{s}$ can be estimated by a minimum norm solution as

$$\hat{\mathbf{s}} = [\mathbf{G}^{AR}]^{+} \cdot \mathbf{v}^{AR} \tag{5}$$

The estimation of $\mathbf{s}$ is based on the fact that activated neural sources are not affected by the particular reference used (Pascual-Marqui and Lehamann, 1993).

Finally, the EEG potentials with IR could be restored approximately by

$$\mathbf{v}^{REST} = \mathbf{G} \cdot \hat{\mathbf{s}} \tag{6}$$

Note that the effectiveness of REST does not depend on the type of inverse solution used. The non-uniqueness of the EEG inverse is not a problem but a way to achieve REST. Generally, the number of equivalent sources is much larger than the number of the scalp electrodes. The unique minimum norm linear inversion is a general choice for this underdetermined problem, and the inversion can be easily conducted by a general pseudo-inverse of the lead field matrix (Yao, 2001).

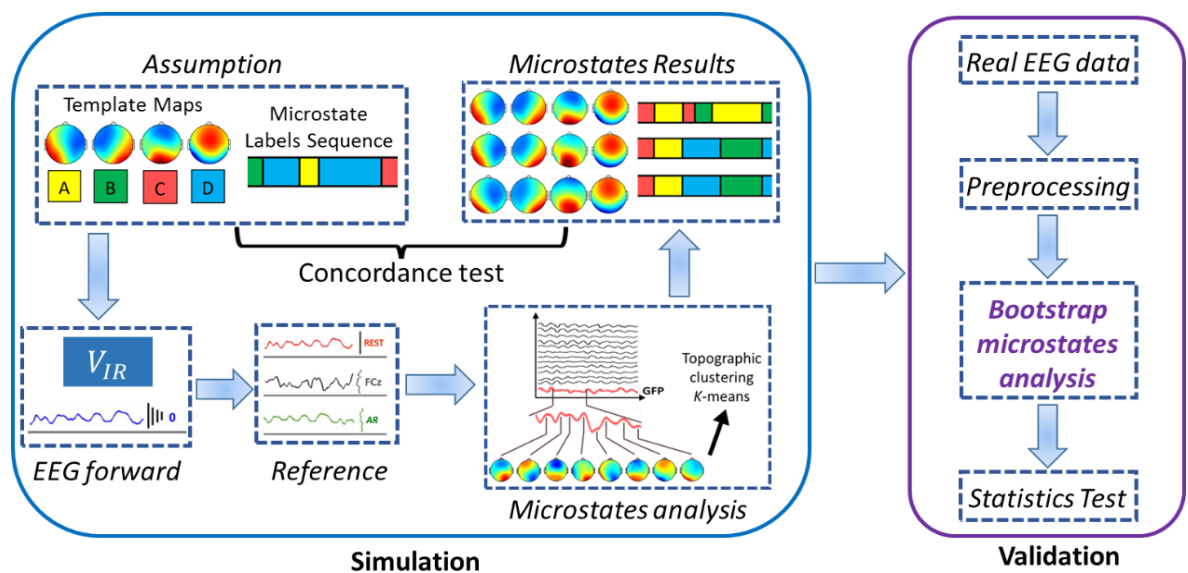

**Fig. 1.** ***The simulation module***: the microstates-based EEG forward model was designed to generate the EEG potentials with IR; after, PBR (FCz electrode), AR, and REST were applied for the reference transforming, respectively; microstates analysis was conducted with the predetermined four types of cluster maps; lastly, the concordance between the microstates results and the assumption was investigated. ***The validation module***: the nonparametric bootstrap model based microstates analysis was applied on the real EEG. statistical analysis was performed across the cluster maps.

## 3. Microstate analysis

The core steps of microstates analysis follow the well-established standard procedures (Lehmann et al., 1987; Strik and Lehmann, 1993; Wackermann et al., 1993). Briefly shown in Fig. 1, firstly, the global field power (GFP) which represents the instantaneous field strength over time is calculated from the multichannel EEG signals; then, peaks of the GFP curve are captured at some time samples where local maxima occur. It is believed that the strongest field strength and the largest signal to noise ratio exist at the GFP peaks (Koenig et al., 2002); afterwards, the electric potentials of all electrodes at the GFP peaks are plotted as topographic maps, due to that high GFP is associated with a stable EEG topography around its peak (Michel, 2009); all the topographic maps are submitted to k-means clustering algorithm to generate a predetermined number of cluster maps; most studies examined on resting state EEG reported the same four archetypal cluster maps as shown in Fig. 1 (Lehmann et al., 2009); finally, all the topographic maps at the GFP peaks will be assigned the labels based on topographic similarity measured by the cosine distance.

### 3.1. Global field power (GFP)

The GFP equals the root mean square across the average referenced electrode values at a given time sample, namely, the standard deviation of all electrodes at a given time (Lehmann et al., 1987; Lehmann and Skrandies, 1980). The usual definition of GFP is based on the EEG potentials referenced to AR (Brunet et al., 2011). Even so, GFP has ever been considered to be independent of the reference choice (Hamburger and v.d. Burgt, 1991; Lehmann et al., 1987; Lehmann and Skrandies, 1980; Murray et al., 2008; Skrandies, 1990). To indicate instantaneous field strength, GFP means how strong the potentials are recorded on average across the electrode montage (Murray et al., 2008; Wackermann et al., 1993). Thus, the other references could be utilized to calculate the GFP as well as AR. For the comparison among different references, the GFP is redefined as following,

$$\text{GFP}_t = \sqrt{\frac{1}{N_e} \sum_{c=1}^{N_e} v_{ct}^2} \tag{7}$$

where $v_{ct}$ is the potential of the $c^{th}$ electrode at a given time sample after the reference being transformed to PBR, AR, REST. This equation has been used in the previous studies (Brodbeck et al., 2012; Hatz et al., 2016; Murray et al., 2008; Wackermann et al., 1993). Hence, the transformation of GFP formulation has no impacts on the GFP computing and the cluster analysis with AR.

## 3.2. Topography similarity

The topographic similarity is measured by the corrected cosine index between two vectors,

$$\text{cci} = \frac{|\mathbf{u} \cdot \mathbf{v}|}{\|\mathbf{u}\|_2 \cdot \|\mathbf{v}\|_2} \tag{8}$$

where $\mathbf{u}$ and $\mathbf{v}$ are the topographic distributions represented by the instantaneous topographic maps or the cluster maps. The instantaneous topographic map is the EEG potentials over all electrodes at a time sample with IR, PBR, AR, or REST. cci ranges from 0 indicating the exactly orthogonal configuration of neural sources, to 1 which means the identical configuration of neural sources regardless of the strength and polarity. Namely, cci describes the extent how the configuration of neural sources is similarly distributed to another one.

If the attribute vectors are normalized by subtracting the means of vector, the index is called centered cosine similarity, and equals to the Pearson correlation coefficient which is the usual measure to evaluate the similarity between two maps (Brandeis et al., 1992). Or say, the cci works equivalently when the Pearson correlation coefficient employs two topographic maps of the EEG potentials referenced with AR.

## 3.3. Clustering algorithm

Compared with the adaptive segmentation method, the clustering algorithm is full of methodological advantages (Lehmann et al., 1987). Two clustering methods used in the microstates analysis are k-means clustering and hierarchical clustering (Pascual-Marqui et al., 1995; Tibshirani and Walther, 2005). However, a study exploring the reliability of the microstates found that the microstates results were highly consistent across two clustering methods (Khanna et al., 2014). Therefore, in this study, only k-means clustering is adopted to study the reference effects to microstates analysis.

The codes of k-means clustering closely matching the results of Cartool - a dedicated software for microstates analysis (Brunet et al., 2011; Michel and Murray, 2012) are shared with us by the Functional Brain Mapping Laboratory in University of Geneva.

## 4. Simulation

The microstates analysis is the identification of the spatiotemporal features of the scalp EEG, thus being in the sensor space rather than the source space. Due to the reference problem and the non-uniqueness of the inverse solution, it is difficult to take one microstate result as the ground truth. An effective way to investigate the effects of EEG references is to generate the microstates projected from the known source activities via simulation.

## 4.1. Microstate-based EEG generation

A microstate-based EEG generative model was proposed according to AR first in (Pascual,1995). This model is expressed as,

$$\begin{cases} \mathbf{V}^m = \mathbf{\Gamma}\mathbf{A} + \mathbf{E}, \; \|\mathbf{\Gamma}_\mu\| = 1 \\ \mathbf{A}(\mu_1,t) \cdot \mathbf{A}(\mu_2,t) = 0, \forall \mu_1 \neq \mu_2 \\ \sum_{k=1}^{N_\mu} \mathbf{A}_{\mu,t}^2 \geq 0 \end{cases} \quad (9)$$

where $\mathbf{V}^m$ is the measured EEG potentials with the size $N_e \times N_t$; $\mathbf{\Gamma}$ is the normalized microstate patterns in the scalp with the size $N_e \times N_\mu$; $\mu$ is the microstate label $1,\cdots,N_\mu$; $\mathbf{A}$ is the intensity of the microstate during the time evolution; $\mathbf{E}$ is the zero mean random noise, independent identically distributed for all time samples. This model assumes that EEG is constituted of $N_\mu$ non-overlapping microstates with a certain topography whose intensity changes over time.

Based on the general EEG forward model (3), we extend the microstates model to the source space so as to guarantee that the simulated EEG is referenced to IR. Considering different levels of the existences of noise, the equation (9) is rewritten as,

$$\mathbf{V}_t^{IR} = \mathbf{G} \cdot (\mathcal{Z}_t \mathbf{A}_t + \mathbf{E}_t^s) + \mathbf{E}_t^m + \mathbf{1} \cdot \mathbf{e}_t, \; \|\mathbf{G}\mathcal{Z}_{\mu,t}\|_2 = 1 \quad (10)$$

where $\mathcal{Z}$ describes the time-varying source patterns, with the size $N_v \times N_\mu \times N_t$; We assume that each source pattern is drawn from a normal distribution with the mean as the template pattern $\mathbf{Z}^{temp}$, $\mathcal{Z}_{\mu,t} \sim N(\mathbf{Z}_\mu^{temp}, \rho \mathbf{I}_{N_v})$, and $\rho$ =0.01; $\mathbf{E}^s$ is the noise in source space; $\mathbf{E}^m$ is the measurement noise; the noise due to the reference electrode is denoted as $\mathbf{e}_t$ which is the smoothed noise based on Gaussian distribution.

By this source model, one can compute the ideal microstate patterns (template maps, i.e. the assumed cluster maps) from $\mathbf{GZ}^{temp}$ and the microstates labels in each time sample from the intensity matrix $\mathbf{A}$. Both the template maps and the microstate label alternating sequence are taken as the ground truth for the comparison of different references in the issue of microstate analysis.

## 4.2. S/N ratio and initial number of clustering centroids

The primal current density of the sources $\mathbf{S}$ and the scalp EEG signals without the measurement noise $\mathbf{V}^p$ can be denoted as

$$\mathbf{S}_t = \mathcal{Z}_t \mathbf{A}_t, \ \mathbf{V}_t^p = \mathbf{G} \cdot (\mathbf{S}_t + \mathbf{E}_t^s) \tag{11}$$

We model the source and measurement noise in the equation (10) as

$$\begin{cases} \text{SNR} = 10\log_{10}\left[\sigma_{signal}^2 / \sigma_{noise}^2\right] \\ \mathbf{E}_t^s \sim N(0, \sigma_s^2 (\mathbf{LL}^T)^{-1}), \ \mathbf{E}_c^m \sim N(0, \sigma_m^2) \end{cases} \tag{12}$$

Here, $\sigma_s^2$ and $\sigma_m^2$ are the variances of noise injected to the signal $\mathbf{S}_t$ in the source space and the signal $\mathbf{V}_c^p$ in the scalp, respectively. The discrete spatial Laplacian matrix $\mathbf{L} = (\mathbf{I}_{N_v} - \frac{1}{6}\mathbf{H}) \otimes \mathbf{I}_3$, and $\mathbf{H}$ has the entries $h_{vv'} = 1$ if $v'$ is a neighbor of source $v$, otherwise 0.

Note that the noise injected in the source space is more likely to change the cluster maps (i.e. the final clustering centroids) than the noise in the sensor space. Due to much noise in the source space, the final clustering centroids may deviate far from the assumed cluster maps. To avoid this, we use more than four initial clustering centroids and eventually only export four centroids which highly correlate to the assumed. Hence, the SNR and the number of the initial clustering centroids are the factors to be investigated in the issue of microstate analysis. Setting the two factors in various levels, the spatial similarity cci is taken as the indicator to evaluate the references.

### 4.3. Evaluation procedures

As the steps in Fig. 1, we generate the resting state EEG with the microstate-based forward model in (10), perform the microstates analysis with different references, and evaluate the cluster maps and the microstate label alternating sequences. This sequel is repeated for different SNRs and number of initial cluster centroids.

The spatial similarity between the generated cluster maps and the assumed is taken as the measure to evaluate to what extent the references affect the spatial distribution of microstates analysis. The ground truth of the microstate label alternating sequences -- the microstate label in each time sample is known in simulation. The topography of the referenced EEG at each time sample was assigned one label based on the spatial similarity between this topography and the generated four cluster maps. The proportion of the correctly identified microstate labels is taken as the measure to reflect how the references affect the dynamical aspect of microstates analysis.

### 4.4. Topography similarity of simulation

Given by 29 channels and 4000 time-samples, we generated four cluster maps and the microstate label on each time sample. By the equation (10), the EEG potentials $\mathbf{v}^{IR}$ were simulated. We firstly subtracted the potentials of FCz from each entry of $\mathbf{v}^{IR}$ as the measured EEG potentials $\mathbf{v}^{PBR}$. Then, the measured EEG potentials was referenced with AR and REST, respectively. In the simulation, the tested SNR is infinity, 3, 2.5, 2, 1.5, 1, and the initial number of clustering centroids is 4, 6, 8, 12, and each combination of the two factors was repetitively ran 30 times. In detail, for each reference, there were totally 4320 cases grouped by the varying factors (6 source SNRs by 6 scalp SNRs by 4 initial numbers of clustering centroids by 30 repetitions) to be investigated.

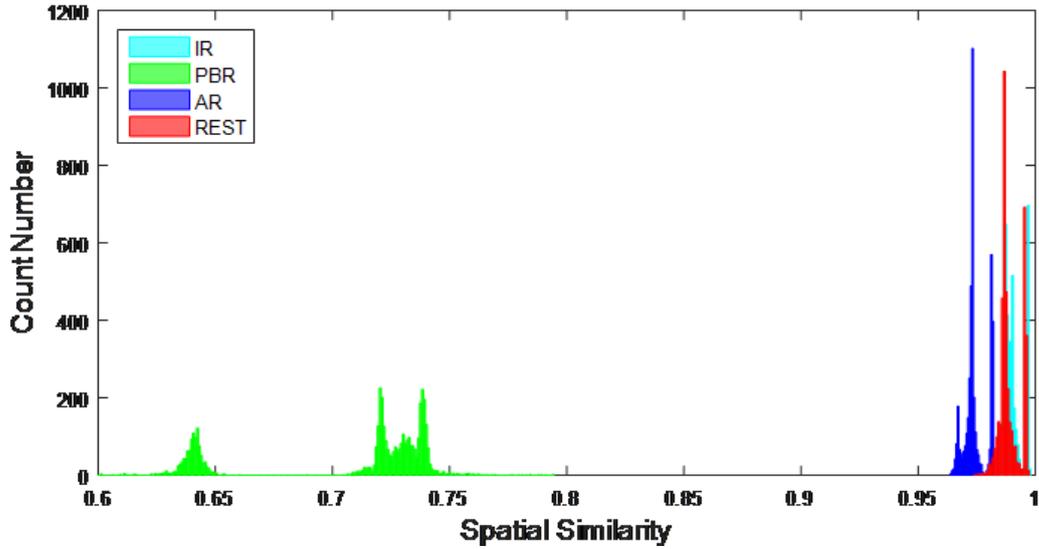

Fig. 2. The histogram of the spatial similarity between the generated and assumed cluster maps by four references. IR: infinity reference; PBR: physical body reference (here, FCz); AR: average reference; REST: reference electrode standardization technique.

Fig. 2 shows the spatial similarity between the assumed and the generated cluster maps by IR, PBR (FCz), AR, and REST, respectively. We depicted all the cases with the same reference as histogram. It is clear in Fig. 2 that PBR is far less likely to recover the assumed cluster maps. Hereafter, PBR is taken no longer into account in the comparison of references. Unsurprisingly, IR gives the highest spatial similarity close to 1 among all the references. Although the spatial similarity by AR is larger than 0.95, REST outperforms AR in all cases.

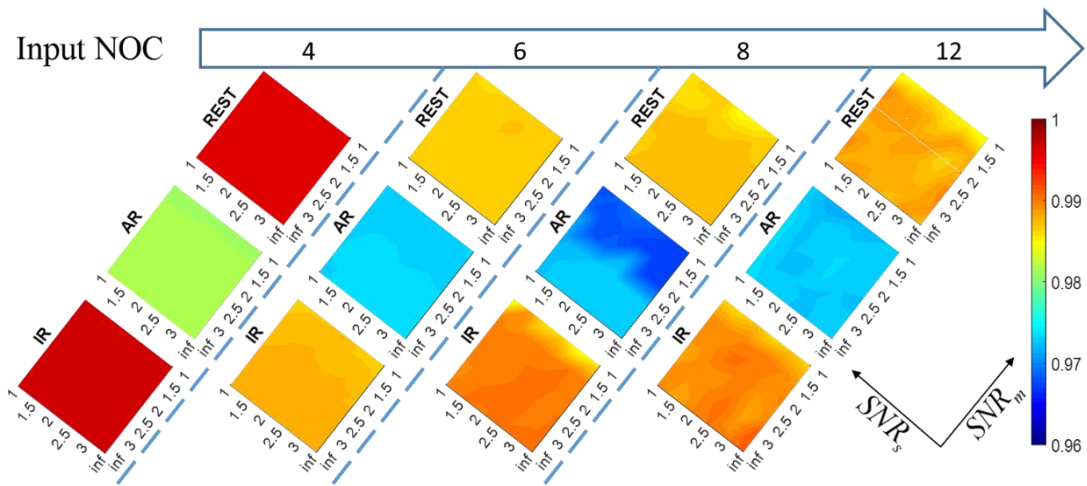

Fig. 3. The spatial similarity with the factors varying. NOC: the initial number of clustering centroids; SNRs: SNR in the source space; SNRm: SNR in the sensor space. The pixel color represents the values of spatial similarity. From left to right, the initial number of clustering centroids increases from 4 to 12. Away from the bottom square toward the top one, three squares are the spatial similarity between the assumption and the cluster map generated by IR, AR, and REST, in order.

The factors influencing the performance of references is the SNR and the initial number of clustering centroids. Fig. 3 shows the performance of references with factors varying. Regardless of how the factors are grouped, REST seems to be similar with IR, while AR always performs worse than REST. The scalp noise affects spatial similarity more greatly than the source noise. This can be seen from the squares where the color changes more along the upper right direction than that in the upper left direction. In addition, the spatial similarity did not have the close relation with the initial number of clustering centroids.

### 4.5. Microstate identification

The most realistic one among all the simulation is selected to present the proportion of the correctly identified microstate labels, as well as the spatial similarity in Fig. 4. The simulated EEG in this case was generated by the factors grouped by the SNR was 1.5 and 1, in the source space and the scalp, respectively, as well as 4 initial clustering centroids. Hoteling's T-squared test between the pairwise references is conducted. Seen from the wide gap and the p-value (<1e-4) between AR and REST, it is evident that REST can give more accurate microstate labels alternating sequences and the cluster maps than AR. The significance level (p=0.4920) between REST and IR indicates that REST has the similar performance to IR, or say, REST can almost identify the correct microstate labels and the cluster maps with IR.

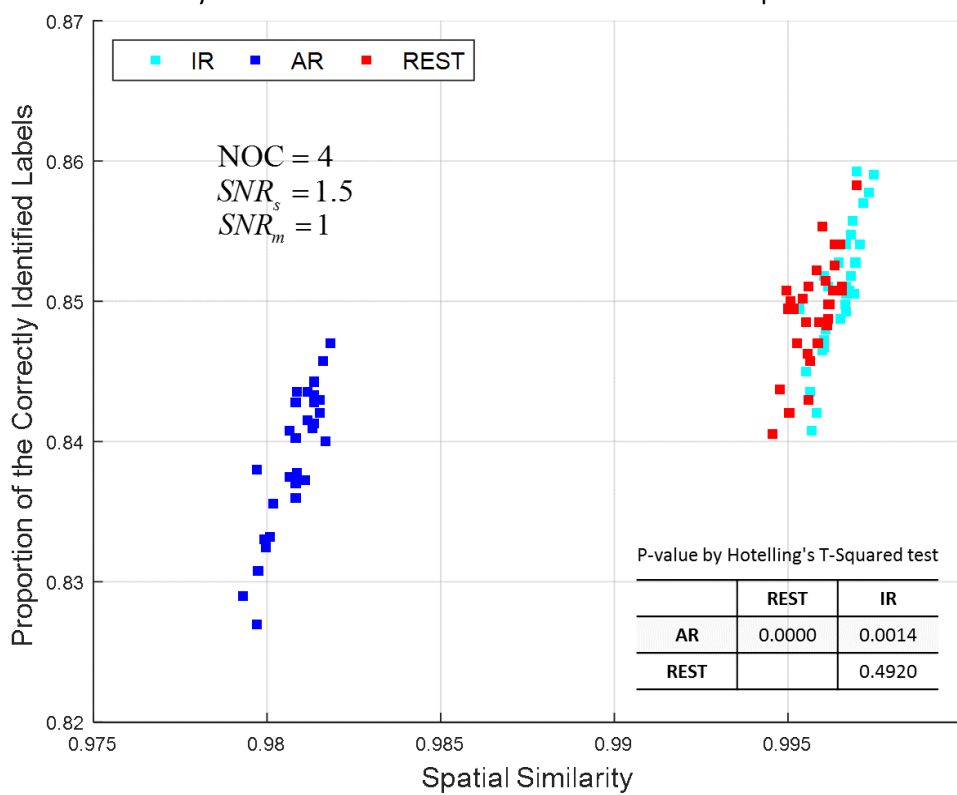

Fig. 4. An illustration of the correctly identified microstate labels. NOC: the initial number of clustering centroids; SNRs: SNR in the source space; SNRm: SNR in the sensor space. The table shows the p-values by Hoteling's T-squared test between the pairwise references.

## 5. Validation by real data

### 5.1. Real EEG data

Real EEG data was acquired from 22 health subjects (age: 24.5±1.3 yrs., gender: 11 males/11 females) who were recruited from University of Electronic Science and Technology of China (UESTC). During the EEG recording, the subject sat on a comfortable armchair and kept relaxed in a salient room with moderate brightness. The subject was instructed to keep eyes open for five minutes and then keep eyes closed for five minutes after a short break. 28 channels (reference electrode FCz and ground electrode AFz excluded) of EEG recordings were sampled at 500 Hz (32bit A/D conversion) within a bandwidth of 0.1-45 Hz by Brain Products System. The study was approved by local ethics committee of UESTC and was conducted according to the given guidelines.

8 seconds segment of a subject with the eyes closed EEG was used to perform microstates analysis in combination with the parametric bootstrap method for the validation. We did the preprocessing such as artifact removal, 1-40 Hz filtering, and reference transforming by AR and REST. It is worthwhile to mention that the FCz is included in the AR transforming. The potential of FCz is not discarded and ascribed a value of 0 for all the time samples. Because the data of FCz is a valid sampled value of the brain electricity, it should be included in the electrode montage and the data analysis (Murray et al., 2008). Adding the PBR electrode FCz, the electrode layout and the number of electrodes are the same with the assumed montage in the simulation.

## 5.2. Parametric Bootstrap model

For the EEG potentials with AR or REST, it can be corresponded to the microstates model in the scalp

$$\mathbf{V}_{(0)} = \mathbf{\Gamma}\mathbf{A} + \mathbf{E}_{(0)} \tag{13}$$

where $\mathbf{V}_{(0)}$ denotes referenced potentials with AR or REST, respectively. $\mathbf{\Gamma}$ is the normalized microstate patterns generated by the k-means clustering. We denote the sum of the noise in source space and scalp as $\mathbf{E}_{(0)}$. The intensity of microstates $\mathbf{A}$, is estimated by minimizing

$$\mathcal{F}(\mathbf{A}_t) = \arg\min \|\mathbf{V}_t - \mathbf{\Gamma}_t \mathbf{A}_t\|_2^2 \tag{14}$$

Based on (13), we reconstructed the noise in each loop of bootstrapped microstates analysis. If $n(n=1,\cdots,20)$ represents the $n^{th}$ bootstrap loop and $\hat{\mathbf{E}}$ is the reconstructed noise, then

$$\hat{\mathbf{E}}_{(n)} \sim N(\mathbf{\mu}_{(n-1)}, \mathbf{\Sigma}_{(n-1)}) \tag{15}$$

Here, $\mathbf{\mu}_{(n-1)}$ and $\mathbf{\Sigma}_{(n-1)}$ are derived from $\mathbf{E}_{(n-1)}$. The mathematical relation could be

$$\begin{cases} \mathbf{\mu}_{(n-1)} = E\{\mathbf{E}_{(n-1)}\}, E\{\mathbf{H}\} = \frac{1}{J}\sum_{j=1}^{J}\mathbf{H}_{i,j}, \mathbf{H} \in R^{I \times J} \\ \mathbf{\Sigma}_{(n-1)} = C\{\mathbf{E}_{(n-1)}(c_1), \mathbf{E}_{(n-1)}(c_2)\}, \forall c_1, c_2 \\ C\{\mathbf{E}_{(n-1)}(c_1), \mathbf{E}_{(n-1)}(c_2)\} = 0, \forall c_1 \neq c_2 \end{cases} \tag{16}$$

Then, we can reconstruct the EEG potentials,

$$\hat{\mathbf{V}}_{(n)} = \mathbf{\Gamma}\mathbf{A} + \hat{\mathbf{E}}_{(n)} \tag{17}$$

The Bootstrap procedure is described in Table 1. Before the reconstruction of the noise in each loop, the reconstructed EEG potentials $\hat{\mathbf{V}}_{(n)}$ was referenced with AR or REST, according to 3) in Table 1.

Table 1. The Parametric Bootstrap Algorithms

---

1) Given $\mathbf{V}^{PBR}$, $\mathbf{G}$, derive $\mathbf{V}_{(0)}^{AR} = (\mathbf{I}_{N_e} - \mathbf{1}\mathbf{f}_1^{\mathbf{T}})\mathbf{V}^{PBR}$, $\mathbf{V}_{(0)}^{REST} = \mathbf{G} \cdot [\mathbf{G}^{AR}]^+ \cdot \mathbf{V}_{AR}$

2) Given $N_\mu$, k-means clustering and equation (9), compute $\mathbf{\Gamma}^{AR}, \mathbf{A}^{AR}, \mathbf{\Gamma}^{REST}, \mathbf{A}^{REST}$

3) Given $n = 1, \cdots, 20$, $\begin{cases} \mathbf{V}_{(n-1)}^{AR} = (\mathbf{I}_{N_e} - \mathbf{1}\mathbf{f}_1^{\mathbf{T}})\hat{\mathbf{V}}_{(n-1)}^{AR}, \mathbf{V}_{(n-1)}^{REST} = \mathbf{G} \cdot [\mathbf{G}^{AR}]^+ \cdot \hat{\mathbf{V}}_{(n-1)}^{REST} \text{ for } n \geq 2 \\ \mathbf{E}_{(n-1)}^{AR} = \mathbf{V}_{(n-1)}^{AR} - \mathbf{\Gamma}^{AR}\mathbf{A}^{AR}, \mathbf{E}_{(n-1)}^{REST} = \mathbf{V}_{(n-1)}^{REST} - \mathbf{\Gamma}^{REST}\mathbf{A}^{REST} \end{cases}$

By equation (10) and (11), compute $\hat{\mathbf{E}}_{(n)}^{AR}$ and $\hat{\mathbf{E}}_{(n)}^{REST}$

By equation (12), compute $\hat{\mathbf{V}}_{(n)}^{AR}$ and $\hat{\mathbf{V}}_{(n)}^{REST}$

By k-means clustering, compute and output $\mathbf{\Gamma}_{(n)}^{AR}$ and $\mathbf{\Gamma}_{(n)}^{REST}$

4) Statistical analysis

---

## 5.3. Results

By the parametric bootstrap method, we conducted the microstates analysis with the reconstructed EEG with 5, 10 and 20 times to obtain ample samples of cluster maps for the statistical analysis. The microstates analysis each time generated four cluster maps labeled as 'A', 'B', 'C' and 'D'. The repeated microstates analysis led to a list of four cluster maps groups. We reorganize the cluster maps into the spatial distribution weight vectors as the following: 1) all the cluster maps are divided into two groups that are 'group AR' and 'group REST', respectively; 2) the cluster maps in each group are reclassified into four sets of cluster maps as the rule of identical labels; 3) since one cluster map is a normalized vector in the multichannel sensor space, the elements in this vector are considered as the weights of corresponding electrodes distributed in the sensor space and together sampled neutral field activities. Taking the weights of an electrode from the set of cluster maps with the identical label forms in one spatial distribution weight vector whose length is the repeated times of microstates analysis. Namely, the repeated times of microstates analysis is the number of samples for each group to perform statistical analysis. Ultimately, paired t-test statistical analysis is performed between two spatial distribution weight vectors of the identical electrode from group AR and group REST, following the order of electrode by electrode and cluster map by cluster map (see Fig. 5). Shown in Fig. 5, there are very high significance levels at 29 electrodes and four labels. Thus, REST generated the significant different cluster maps from AR. The vertical axis is the p-value after paired t-test between two spatial distribution weight vectors from group AR and group REST. The green line is the threshold of significance level ($p = 0.05$). The scatter points with green, red and cyan color are the statistical result by repeating the bootstrapped microstates analysis with 5, 10 and 20 times, respectively. Since most p-values for 20 times microstates analysis are zero, we show them plus 1e-20. All the scatter points are under the green line with $p \ll 0.05$.

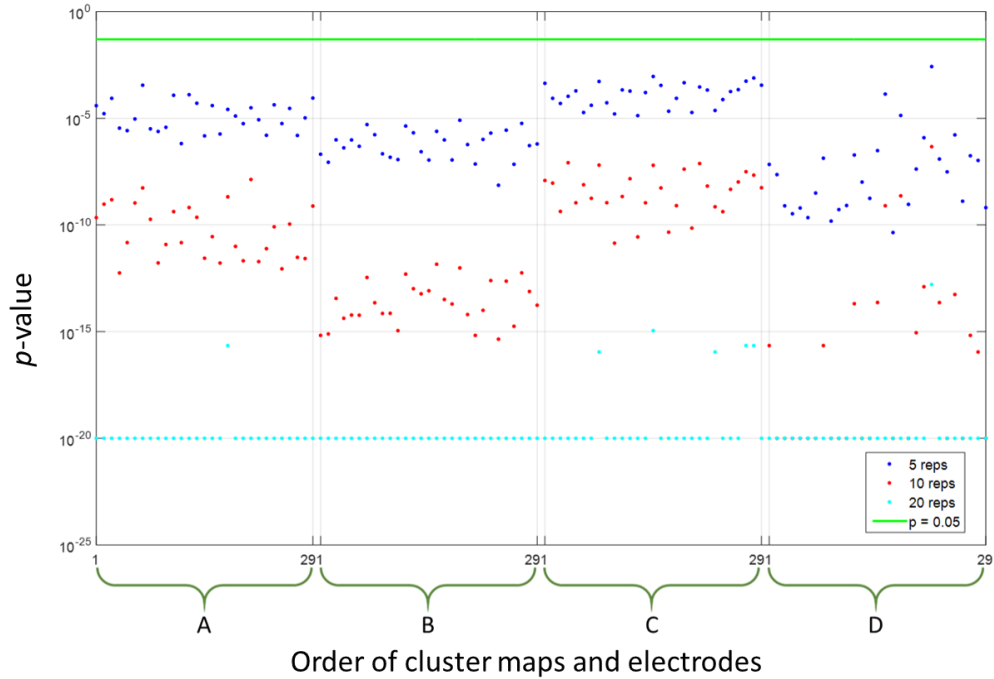

fig. 5. The p-value between AR and REST after bootstrapped microstates analysis as to the weights over electrodes in the cluster map. The horizontal axis shows how the cluster maps and electrodes are ordered in this plot. In the identical order of electrodes, the cluster maps from left to right are labelled as 'A', 'B', 'C' and 'D', respectively.

## 6. Discussion

The present study confirms that the microstates analysis is affected by the references. Fig. 2 indicates that PBR is not reliable for microstates analysis. REST achieves the closer microstate identification to the assumed than AR, according to Fig. 3 and Fig. 4. And the superior performance of REST is not affected by the SNR and the initial number of clustering centroids. The microstates analysis on the real EEG data shows that REST generates four significantly different cluster maps from AR in Fig. 5. Since the microstate label of the original map in each time sample comes from the spatial similarity between the original map itself and the cluster maps, it is natural to infer that the microstate label alternating sequence on the real EEG data may be remarkably different between REST and AR. One may wonder if the GFP results into dissimilar cluster maps for different references. Note that the aim of GFP is to extract the maps at the instants of its peaks for the clustering analysis. The instants of the peaks due to different references will influence the cluster maps, especially for the microstate label alternating series (Khanna et al., 2015).

The microstates topographies (cluster maps) are not directly picked from the distinct topographic maps but affected by the clustering process. In each iteration of k-means clustering algorithm, the k template maps (i.e. clustering centroids) are updated by averaging all the topographic maps which belong to the same class (Murray et al., 2008; Pascual-Marqui et al., 1995). Until convergence, the cluster maps are generated as the microstates topographies. Taking averaging in each iteration has the effect to dilute the original small topography difference. Therefore, the spatial similarity of cluster maps shown in Fig. 2 and Fig. 3 do not manifest rather large difference between AR and REST.

Using the clustering methods, microstates analysis decomposes the multichannel EEG into several spatial maps alternating the stationary states rather than repetitive spatiotemporal (channel by time) patterns analysis (Takeda et al., 2016). This is to say that microstates analysis is more related to the spatial maps than the consecutive spatial patterns in the temporal domain. Due to this, it was of less ample evidences that REST outperformed AR to a very large extent in microstates analysis, even though REST could be especially efficient for identifying and recovering the temporal information of EEG recordings (Yao, 2001; Zhai and Yao, 2004).

The uncontrolled temporal dynamical bias of the reference electrode injected into the EEG recordings will eventually affect the microstates analysis. REST has been applied to correct the temporal biases, such as reducing the systematic shifts in the distribution of the frequency power (Yao et al., 2005), estimating the objective coherency maps and the functional connectivity (Chella et al., 2016; Marzetti et al., 2007), recovering the network configuration exactly (Qin et al., 2010); showing the unbiased audiovisual effects (Tian and Yao, 2013). Our findings in this study validates the reference effects in microstates analysis and indicates that REST could correct the spatiotemporal bias.

The reliability of AR needs further investigation. Its advantage is that EEG potentials nearly sum to zero if the head is modelled as a closed sphere, requiring a large number of electrodes with a whole dense distribution on the head (Offner, 1950). However, the requirements are hard to meet in the practice: 1) one constraint comes from the head model-lacking some of the anatomical and physiological properties of the head. The head is not the perfect closed sphere, with some openings (e.g. ears) and the conductivity of the brain tissues and skull is anisotropic. 2) the other constraint is the limited electrode density and incomplete electrode coverage in the EEG recording. The experimenters usually place the electrodes on the upper half of the head, neglecting the lower part (e.g. face). Collectively, the practical reasons result in average potential differing from zero, preventing AR to be considered as an ideal reference (Hu et al., 2017; Yao, 2017).

It is important to remind that we had not conducted any study on the clustering algorithm and applied REST on the clinical psychiatric EEG for microstates analysis. This study was focused on investigating the effects of the references on the cluster maps and microstate label alternating series reflecting the dynamical information. In the future, we hope to explore the performance of REST with real patient EEG data in the microstates analysis.

## 7. Conclusion

In this study, we investigated the effects of the references on the microstates analysis by the simulated and real EEG. PBR is not suited for microstates analysis and REST always outperform AR to some extent. REST could generate the highly similar microstates features to the assumption, and produce significantly different cluster maps from AR by the real EEG data. To sum up, microstates analysis is affected by the references.

## Acknowledgement

The authors declare no conflict of interest. This work was co-funded by NSFC #61673090, #81330032 and 111 project B12027. We would like to thank Anna Custo and Frédéric Grouiller for providing the clustering algorithm.

## Reference

Bertrand, O., Perrin, F., and Pernier, J. (1985). A theoretical justification of the average reference in topographic evoked potential studies. *Electroencephalogr. Clin. Neurophysiol. Potentials Sect.* 62, 462–464. doi:10.1016/0168-5597(85)90058-9.

Brandeis, D., Naylor, H., Halliday, R., Callaway, E., and Yano, L. (1992). Scopolamine Effects on Visual Information Processing, Attention, and Event-Related Potential Map Latencies. *Psychophysiology* 29,

315–335. doi:10.1111/j.1469-8986.1992.tb01706.x.

Brodbeck, V., Kuhn, A., von Wegner, F., Morzelewski, A., Tagliazucchi, E., Borisov, S., et al. (2012). EEG microstates of wakefulness and NREM sleep. *Neuroimage* 62, 2129–2139. doi:10.1016/j.neuroimage.2012.05.060.

Brunet, D., Murray, M. M., and Michel, C. M. (2011). Spatiotemporal Analysis of Multichannel EEG: CARTOOL. *Comput. Intell. Neurosci.* 2011, 1–15. doi:10.1155/2011/813870.

Chella, F., Pizzella, V., Zappasodi, F., and Marzetti, L. (2016). Impact of the reference choice on scalp EEG connectivity estimation. *J. Neural Eng.* 13, 36016. doi:10.1088/1741-2560/13/3/036016.

Christodoulakis, M., Hadjipapas, A., Papathanasiou, E. S., Anastasiadou, M., Papacostas, S. S., and Mitsis, G. D. (2013). "On the Effect of Volume Conduction on Graph Theoretic Measures of Brain Networks in Epilepsy," in *Modern Electroencephalographic Assessment Techniques: Theory and Applications*, ed. V. Sakkalis (New York, NY: Springer New York), 103–130. doi:10.1007/7657_2013_65.

Dien, J. (1998). Issues in the application of the average reference: Review, critiques, and recommendations. *Behav. Res. Methods, Instruments, Comput.* 30, 34–43. doi:10.3758/BF03209414.

Garneski, T. M., and Steelman, H. F. (1958). Equalizing ear reference resistance in monopolar recording to eliminate artifactual temporal lobe asymmetry. *Electroencephalogr. Clin. Neurophysiol.* 10, 736–738. doi:10.1016/0013-4694(58)90081-6.

Geselowitz, D. B. (1998). The zero of potential. *IEEE Eng. Med. Biol. Mag.* 17, 128–132. doi:10.1109/51.646230.

Hamburger, H. L., and v.d. Burgt, M. A. G. (1991). Global Field Power measurement versus classical method in the determination of the latency of evoked potential components. *Brain Topogr.* 3, 391–396. doi:10.1007/BF01129642.

Hatz, F., Hardmeier, M., Bousleiman, H., Rüegg, S., Schindler, C., and Fuhr, P. (2016). Reliability of functional connectivity of EEG applying microstates-segmented versus classical calculation of phase lag index. *Brain Connect.*, 1–29. doi:10.1089/brain.2015.0368.

Hu, S., Lai, Y., Valdés-Sosa, P. A., Bringas-Vega, M. L., and Yao, D. (2017). How do reference montage and electrodes setup affect the measured scalp EEG potentials? *J. Neural Eng.* 22, 56. doi:10.1088/1741-2552/aaa13f.

Hu, S., Yao, D., and Valdes-Sosa, P. A. (2018). Unified Bayesian estimator of EEG reference at infinity: rREST. Available at: http://arxiv.org/abs/1802.02268.

Kaiser, D. A. (2000). QEEG:State of the Art, or State of Confusion. *J. Neurother.* 4, 57–75. doi:10.1300/J184v04n02_07.

Khanna, A., Pascual-Leone, A., and Farzan, F. (2014). Reliability of Resting-State Microstate Features in Electroencephalography. *PLoS One* 9, e114163. doi:10.1371/journal.pone.0114163.

Khanna, A., Pascual-Leone, A., Michel, C. M., and Farzan, F. (2015). Microstates in resting-state EEG: Current status and future directions. *Neurosci. Biobehav. Rev.* 49, 105–113. doi:10.1016/j.neubiorev.2014.12.010.

Kikuchi, M., Koenig, T., Munesue, T., Hanaoka, A., Strik, W., Dierks, T., et al. (2011). EEG Microstate Analysis in Drug-Naive Patients with Panic Disorder. *PLoS One* 6, e22912. doi:10.1371/journal.pone.0022912.

Koenig, T., Hubl, D., and Mueller, T. J. (2002). Decomposing the EEG in time, space and frequency: a formal model, existing methods, and new proposals. *Int Congr Ser* 1232, 317–321. doi:10.1016/S0531-5131(01)00724-5.

Lehmann, D. (1971). Multichannel topography of human alpha EEG fields. *Electroencephalogr. Clin. Neurophysiol.* 31, 439–449. doi:10.1016/0013-4694(71)90165-9.

Lehmann, D., Faber, P. L., Galderisi, S., Herrmann, W. M., Kinoshita, T., Koukkou, M., et al. (2005). EEG

microstate duration and syntax in acute, medication-naïve, first-episode schizophrenia: A multi-center study. *Psychiatry Res. - Neuroimaging* 138, 141–156. doi:10.1016/j.pscychresns.2004.05.007.

Lehmann, D., Ozaki, H., and Pal, I. (1987). EEG alpha map series: brain micro-states by space-oriented adaptive segmentation. *Electroencephalogr. Clin. Neurophysiol.* 67, 271–288. doi:10.1016/0013-4694(87)90025-3.

Lehmann, D., Pascual-Marqui, R., and Michel, C. (2009). EEG microstates. *Scholarpedia* 4, 7632. doi:10.4249/scholarpedia.7632.

Lehmann, D., and Skrandies, W. (1980). Reference-free identification of components of checkerboard-evoked multichannel potential fields. *Electroencephalogr. Clin. Neurophysiol.* 48, 609–621. doi:10.1016/0013-4694(80)90419-8.

Lehmann, D., Strik, W. K., Henggeler, B., Koenig, T., and Koukkou, M. (1998). Brain electric microstates and momentary conscious mind states as building blocks of spontaneous thinking: I. Visual imagery and abstract thoughts. *Int. J. Psychophysiol.* 29, 1–11. doi:10.1016/S0167-8760(97)00098-6.

Liu, Q., Balsters, J. H., Baechinger, M., Van Der Groen, O., Wenderoth, N., and Mantini, D. (2015). Estimating a neutral reference for electroencephalographic recordings: the importance of using a high-density montage and a realistic head model. *J. Neural Eng.* 12, 56012. doi:10.1088/1741-2560/12/5/056012.

Marzetti, L., Nolte, G., Perrucci, M. G., Romani, G. L., and Del Gratta, C. (2007). The use of standardized infinity reference in EEG coherency studies. *Neuroimage* 36, 48–63. doi:10.1016/j.neuroimage.2007.02.034.

McAvoy, M., and Little, S. C. (1949). Technical observations on the use of independent ear electrodes. *Dis. Nerv. Syst.* 10, 207–10. Available at: http://www.ncbi.nlm.nih.gov/pubmed/18132841.

Michel, C. M. (2009). *Electrical neuroimaging*. Cambridge University Press.

Michel, C. M., and Koenig, T. (2017). EEG microstates as a tool for studying the temporal dynamics of whole-brain neuronal networks: A review. *Neuroimage*, 1–21. doi:10.1016/j.neuroimage.2017.11.062.

Michel, C. M., and Murray, M. M. (2012). Towards the utilization of EEG as a brain imaging tool. *Neuroimage* 61, 371–385. doi:10.1016/j.neuroimage.2011.12.039.

Michel, C. M., Murray, M. M., Lantz, G., Gonzalez, S., Spinelli, L., and Grave de Peralta, R. (2004). EEG source imaging. *Clin. Neurophysiol.* 115, 2195–2222. doi:10.1016/j.clinph.2004.06.001.

Murray, M. M., Brunet, D., and Michel, C. M. (2008). Topographic ERP analyses: a step-by-step tutorial review. *Brain Topogr.* 20, 249–64. doi:10.1007/s10548-008-0054-5.

Nishida, K., Morishima, Y., Yoshimura, M., Isotani, T., Irisawa, S., Jann, K., et al. (2013). EEG microstates associated with salience and frontoparietal networks in frontotemporal dementia, schizophrenia and Alzheimer's disease. *Clin. Neurophysiol.* 124, 1106–1114. doi:10.1016/j.clinph.2013.01.005.

Nunez, P. L. (1991). The linked-reference issue in EEG and ERP recording": Comments on the paper by Miller, Lutzenberger and Elbert. *J. Psychophysiol.* 5, 279–280.

Nunez, P. L., and Srinivasan, R. (2006). *Electric Fields of the Brain*. The 2nd ed. Oxford University Press doi:10.1093/acprof:oso/9780195050387.001.0001.

Nunez, P. L., Srinivasan, R., Westdorp, A. F., Wijesinghe, R. S., Tucker, D. M., Silberstein, R. B., et al. (1997). EEG coherency I: Statistics, reference electrode, volume conduction, Laplacians, cortical imaging, and interpretation at multiple scales. *Electroencephalogr. Clin. Neurophysiol.* 103, 499–515. doi:10.1016/S0013-4694(97)00066-7.

Offner, F. F. (1950). The EEG as potential mapping: The value of the average monopolar reference. *Electroencephalogr. Clin. Neurophysiol.* 2, 213–214. doi:10.1016/0013-4694(50)90040-X.

Pascual-Marqui, R. D., and Lehamann, D. (1993). Topographic maps, source localization inference, and the reference electrode: comments on a paper by Desmedt et al. *Electroencephalogr. Clin. Neurophysiol. Potentials Sect.* 88, 532–6.

Pascual-Marqui, R. D., Michel, C. M., and Lehmann, D. (1995). Segmentation of Brain Electrical Activity into Microstates; Model Estimation and Validation. *IEEE Trans. Biomed. Eng.* 42, 658–665. doi:10.1109/10.391164.

Qin, Y., Xu, P., and Yao, D. (2010). A comparative study of different references for EEG default mode network: The use of the infinity reference. *Clin. Neurophysiol.* 121, 1981–1991. doi:10.1016/j.clinph.2010.03.056.

Shaw, J. C. (1984). Correlation and coherence analysis of the EEG: A selective tutorial review. *Int. J. Psychophysiol.* 1, 255–266. doi:10.1016/0167-8760(84)90045-X.

Skrandies, W. (1990). Global field power and topographic similarity. *Brain Topogr.* 3, 137–141. doi:10.1007/BF01128870.

Strik, W. K., and Lehmann, D. (1993). Data-determined window size and space-oriented segmentation of spontaneous EEG map series. *Electroencephalogr. Clin. Neurophysiol.* 87, 169–174. doi:10.1016/0013-4694(93)90016-O.

Takeda, Y., Hiroe, N., Yamashita, O., and Sato, M. (2016). Estimating repetitive spatiotemporal patterns from resting-state brain activity data. *Neuroimage* 133, 251–265. doi:10.1016/j.neuroimage.2016.03.014.

Teplan, M. (2002). Fundamentals of EEG measurement. *Meas. Sci. Rev.* 2, 1–11.

Tian, Y., and Yao, D. (2013). Why do we need to use a zero reference? Reference influences on the ERPs of audiovisual effects. *Psychophysiology* 50, 1282–1290. doi:10.1111/psyp.12130.

Tibshirani, R., and Walther, G. (2005). Cluster Validation by Prediction Strength. *J. Comput. Graph. Stat.* 14, 511–528. doi:10.1198/106186005X59243.

Travis, F. (1994). A Second Linked-Reference Issue: Possible Biasing of Power and Coherence Spectra. *Int. J. Neurosci.* 75, 111–117. doi:10.3109/00207459408986294.

Wackermann, J., Lehmann, D., Michel, C. M., and Strik, W. K. (1993). Adaptive segmentation of spontaneous EEG map series into spatially defined microstates. *Int. J. Psychophysiol.* 14, 269–283. doi:10.1016/0167-8760(93)90041-M.

Yao, D. (2001). A method to standardize a reference of scalp EEG recordings to a point at infinity. *Physiol. Meas.* 22, 693–711. doi:10.1088/0967-3334/22/4/305.

Yao, D. (2017). Is the Surface Potential Integral of a Dipole in a Volume Conductor Always Zero? A Cloud Over the Average Reference of EEG and ERP. *Brain Topogr.* 30, 161–171. doi:10.1007/s10548-016-0543-x.

Yao, D., Wang, L., Arendt-Nielsen, L., and N Chen, A. C. (2007). The effect of reference choices on the spatio-temporal analysis of brain evoked potentials: The use of infinite reference. *Comput. Biol. Med.* 37, 1529–1538. doi:10.1016/j.compbiomed.2007.02.002.

Yao, D., Wang, L., Oostenveld, R., Nielsen, K. D., Arendt-Nielsen, L., and Chen, A. C. N. (2005). A comparative study of different references for EEG spectral mapping: the issue of the neutral reference and the use of the infinity reference. *Physiol. Meas.* 26, 173–184. doi:10.1088/0967-3334/26/3/003.

Zhai, Y., and Yao, D. (2004). A study on the reference electrode standardization technique for a realistic head model. *Comput. Methods Programs Biomed.* 76, 229–238. doi:10.1016/j.cmpb.2004.07.002.